\documentclass[preprint2,trackchanges]{aastex6}

\usepackage{amsmath,multirow,rotating}
\usepackage[graphicx]{realboxes}
\hypersetup{breaklinks=true,colorlinks=true,citecolor=blue,linkcolor=blue,urlcolor=blue}

\shorttitle{SN Rates in LOSS, II}
\shortauthors{Graur et al.}

\begin{document}

\title{LOSS Revisited --- II: The Relative Rates of Different Types of Supernovae Vary between Low- and High-mass Galaxies}

\author{Or Graur\altaffilmark{1}}
\affil{Harvard-Smithsonian Center for Astrophysics, 60 Garden St., Cambridge, MA 02138, USA; \url{or.graur@cfa.harvard.edu}}
\affil{CCPP, New York University, 4 Washington Place, New York, NY 10003, USA}
\affil{Department of Astrophysics, American Museum of Natural History, New York, NY 10024, USA}

\author{Federica B. Bianco}
\affil{CCPP, New York University, 4 Washington Place, New York, NY 10003, USA}
\affil{Center for Urban Science and Progress, New York University, 1 MetroTech Center, Brooklyn, NY 11201, USA}

\author{Maryam Modjaz}
\affil{CCPP, New York University, 4 Washington Place, New York, NY 10003, USA}

\author{Isaac Shivvers, Alexei V. Filippenko, and Weidong Li\altaffilmark{2}}
\affil{Department of Astronomy, University of California, Berkeley, CA 94720-3411, USA}

\author{Nathan Smith}
\affil{Steward Observatory, University of Arizona, 933 North Cherry Avenue, Tucson, AZ 85721, USA}

\altaffiltext{1}{NSF Astronomy and Astrophysics Postdoctoral Fellow.}
\altaffiltext{2}{Deceased 2011 December 12.}


\begin{abstract}
\noindent In Paper I of this series, we showed that the ratio between stripped-envelope (SE) supernova (SN) and Type II SN rates reveals a significant SE SN deficiency in galaxies with stellar masses $\lesssim 10^{10}~{\rm M}_\sun$. Here, we test this result by splitting the volume-limited subsample of the Lick Observatory Supernova Search (LOSS) SN sample into low- and high-mass galaxies and comparing the relative rates of various SN types found in them. The LOSS volume-limited sample contains 180 SNe and SN impostors and is complete for SNe Ia out to 80 Mpc and core-collapse SNe out to 60 Mpc. All of these transients were recently reclassified by us in \citet{2016arXiv160902922S}. We find that the relative rates of some types of SNe differ between low- and high-mass galaxies: SNe Ib and Ic are underrepresented by a factor of $\sim3$ in low-mass galaxies. These galaxies also contain the only examples of SN 1987A-like SNe in the sample and host about $9$ times as many SN impostors.  Normal SNe Ia seem to be $\sim 30$\% more common in low-mass galaxies, making these galaxies better sources for homogeneous SN Ia cosmology samples. The relative rates of SNe IIb are consistent in both low- and high-mass galaxies. The same is true for broad-line SNe~Ic, although our sample includes only two such objects. The results presented here are in tension with a similar analysis from the Palomar Transient Factory, especially as regards SNe IIb.
\end{abstract}



\section{Introduction}
\label{sec:intro}

This is the second paper in a series that reanalyzes the supernova (SN) sample assembled by the Lick Observatory Supernova Search (LOSS; \citealt{2000AIPC..522..103L,2001ASPC..246..121F,2003fthp.conf..171F,2005ASPC..332...33F}) in order to constrain SN progenitor models. In Paper I \citep{2016arXiv160902921G}, we remeasured the LOSS SN rates (first measured by \citealt{li2011rates}) and found that the ratio between the rates of stripped-envelope supernovae (SE~SNe; i.e., SNe~IIb, Ib, Ic, broad-lined Ic or Ic-BL, and peculiar examples of these subtypes; see \citealt{1997ARA&A..35..309F} for a review) and SNe II (i.e., SNe IIP, IIL, IIn, and peculiar examples of these subtypes) was smaller, by a factor of $\sim 3$, in galaxies with stellar masses $\lesssim 10^{10}~{\rm M}_\sun$ than in more massive galaxies.

The SN rates in Paper I can be regarded as ``absolute'' rates---they measured how many SNe, of a given type, explode per unit time per unit mass. We measured these rates for three broad SN categories: SNe~Ia, SNe~II, and SE~SNe. In this paper, we use a subsample of the LOSS SN sample to measure the fractions of different SN subtypes within this sample. These fractions can be thought of as ``relative'' rates---they measure which fraction of all SNe that explode in nature are of a given subtype. If relative rates are measured from a SN sample that is complete within a given volume (i.e., ``volume-limited''), and the host galaxies of these SNe are representative of the galaxy luminosity function within that volume, the relative SN rates should correlate with the relative rates of their respective progenitor stars. Throughout this work, we use the terms ``relative rates'' and fractions interchangeably. Owing to the careful way in which the LOSS volume-limited subsample was constructed (see below), the relative rates we measure from it allow us to go a step further than in Paper I and study different subtypes of SNe (e.g., SNe~IIb, Ib, and Ic) in detail.

The subsample of the full LOSS SN sample we used here was constructed by \citet[hereafter L11]{li2011LF} to measure SN luminosity functions and relative rates. This sample is complete to all core-collapse (CC) SNe out to 60 Mpc and SNe Ia out to 80 Mpc (and hence it is volume limited). We describe this sample in detail in Section~\ref{sec:galaxies}. Recently, \citet{2016arXiv160902922S} reclassified the SNe in this sample and found that SNe Ib, which L11 initially suggested were roughly half as common as SNe Ic, are actually $1.7 \pm 0.9$ times as common.

\citet{2011MNRAS.412.1522S} used the L11 sample to argue that for a standard initial mass function (IMF), Wolf--Rayet stars could account for only half of the observed fractions of SE SNe. However, in that work, the authors took a conservative tack and treated the LOSS SN relative rates as monolithic, in the sense that the same rates applied to all types of galaxies. In Paper I we showed that galaxies with stellar masses lower than $\sim10^{10}~{\rm M}_\sun$ were less efficient at producing SE SNe than more massive galaxies. In Section~\ref{sec:vol_lim}, we split the LOSS volume-limited sample according to this mass criterion to test whether the same trend is evident in the relative SN rates. If so, we expect to see a lower fraction of SE SNe in galaxies with $M_\star \lesssim 10^{10}~{\rm M}_\sun$. 

In Paper I we calculated the rates of SNe Ia, SE SNe, and SNe II. Here, we further subdivide these SN types into various subtypes (e.g., SE SNe into SNe Ib, Ic, Ic-BL, IIb, etc.) and measure the relative rates of each subtype. We further compare the relative rates of each subtype in low- and high-mass galaxies, using different mass cuts.

We find that SNe Ib and Ic are underrepresented in low-mass galaxies by a factor of $\sim 3$. On the other hand, the relative rates of SNe IIb are consistent between low-mass and high-mass galaxies. There are about $9$ times more SN impostors in low-mass galaxies than in high-mass galaxies (but this might be due to a selection effect). The low-mass galaxies also host the only SN 1987A-like SNe in the LOSS volume-limited sample. Importantly for cosmology, normal SNe Ia are more common in low-mass galaxies.

These results are in tension with those of the Palomar Transient Factory (PTF; \citealt{2009PASP..121.1334R}), as presented by \citet{2010ApJ...721..777A} and \citet{2012IAUS..279...34A}. We discuss the differences between the LOSS and PTF results---and their SN and galaxy samples---in Section~\ref{subsec:vol_PTF}.

Although \citet{smartt2009mnras} also measured SN relative rates, we do not compare the results shown here to theirs. First, their sample consisted of all SNe whose discovery was made public, through International Astronomical Union Circulars, between 1998 January 1 and 2008 June 30 out to 28 Mpc. As these SNe were discovered by different surveys, the completeness of this sample is unknown. Second, \citet{smartt2009mnras} did not split their sample according to either the mass (as done here) or the luminosity (as in \citealt{2010ApJ...721..777A}) of their SN host galaxies. For a discussion of the differences between the overall LOSS relative rates from L11 and the rates reported by \citet{smartt2009mnras}, see \citet{2011MNRAS.412.1522S}.

Relative SN rates measured from a complete sample can also be compared to absolute SN rates from volumetric surveys of the local Universe (e.g., \citealt{li2011rates} and our Paper I). \citet{1997A&A...322..431C} combined SN samples from five different surveys and measured SN Ia, SE SN, and SN~II rates in galaxies of different morphological types. However, owing to the small number of SE SNe in their sample, it is hard to draw any conclusions regarding this family of SNe, and we do not compare our results to theirs.

We summarize our results in Section~\ref{sec:discuss}.


\section{The LOSS Volume-limited Sample}
\label{sec:galaxies}

For a full description of the LOSS sample, we refer the reader to Paper I and to \citet{2011MNRAS.412.1419L}. The LOSS volume-limited sample is ``volume-limited'' in the sense that the authors assume that, given their detection efficiency, they did not miss any SNe that exploded within the galaxies targeted by LOSS out to 80 Mpc for SNe Ia and 60 Mpc for CC SNe. As shown in Figure 4 of L11 and discussed in their section~2.5, LOSS achieved a completeness of $>98$\% for SNe Ia within 80 Mpc, but only $>80$\% for CC~SNe within 60 Mpc. To overcome this problem, L11 measured a completeness correction for each SN in the sample based on its light curve and their detection efficiency. 

Although the LOSS volume-limited sample is complete to all SNe within 60 Mpc, the galaxy sample is not an accurate representation of the galaxy luminosity distribution within the volume it probes. As Figure 1 from \citet{2011MNRAS.412.1419L} shows, both the full LOSS sample and the volume-limited subsample are complete down to galaxies with $K$-band absolute magnitudes of $\sim -24$ mag, but deficient in lower-luminosity galaxies. This deficiency, which is also apparent in the galaxy distributions shown in the various rate figures in Paper I, is caused by the targeted nature of the LOSS survey. In Section~\ref{subsec:vol_caveats}, we address how this incompleteness might affect the rates measured here.

\citet{2016arXiv160902922S} reclassified the L11 sample using all currently available spectra, newly developed tools, and the knowledge that has accrued about SN classification since the original publication of L11; see their Table 1 for a detailed summary of the new classifications. The volume-limited sample used here contains 180 SNe and SN impostors comprising 74 SNe Ia, 34 SE SNe, 65 SNe II, and seven SN impostors. 

Here, we regard as SN impostors those SN candidates that may initially appear to spectroscopically resemble SNe~IIn, but are less luminous than SNe~IIn and show photometric variability before (and sometimes after) the explosion on timescales of days to years, indicating that these may not be terminal explosions but rather eruptions. Based on these criteria, L11 classified SNe 1999bw, 2000ch, 2001ac, 2002kg, and 2003gm as SN impostors. \citet{2011MNRAS.415..773S} later changed the original classifications of SNe 2002bu and 2006bv from SNe IIn to SN impostors. We refer the reader to \citet{2011MNRAS.415..773S} and \citet{2014ARA&A..52..487S} for further discussion of this diverse class of transients.

Our definition of SE~SNe has changed slightly between Paper I and this work. In Paper I, where we used the original control times from \citet{li2011rates} to measure SN rates, we followed \citet{li2011rates} and did not include SNe~IIb in the SE SN category. Here, this is no longer necessary, and SE~SNe refer to all SNe~IIb, Ib, Ic, Ic-BL, ``Ca-rich,''\footnote{SNe 2003H, 2003dr, and 2005E; \citet{2010Natur.465..322P}.} and peculiar SE~SNe. Likewise, SNe II refer to all SNe IIP, IIL, IIn, SN 1987A-like SNe, and peculiar SNe II. As in Paper I, we group SNe~IIP and IIL into one subtype. We use the same definition for SNe~Ia as in \citet{2016arXiv160902922S}, which includes all normal SNe~Ia, subluminous SN 1991bg-like SNe~Ia (e.g., \citealt{1992AJ....104.1543F}), overluminous SN 1991T-like (e.g., \citealt{1992ApJ...384L..15F}), and SN 1999aa-like SNe~Ia (which fall between normal SNe~Ia and SN 1991T-like SNe~Ia; \citealt{2004AJ....128..387G}), SN 2002cx-like SNe~Ia (\citealt{2003PASP..115..453L,2006AJ....132..189J}; also called SNe Iax---\citealt{2013ApJ...767...57F}), and SN 2002es-like SNe~Ia (which have shared properties with both SN 1991bg-like and SN-2002cx-like SNe~Ia; \citealt{2012ApJ...751..142G}). 

\begin{figure}
 \includegraphics[width=0.47\textwidth]{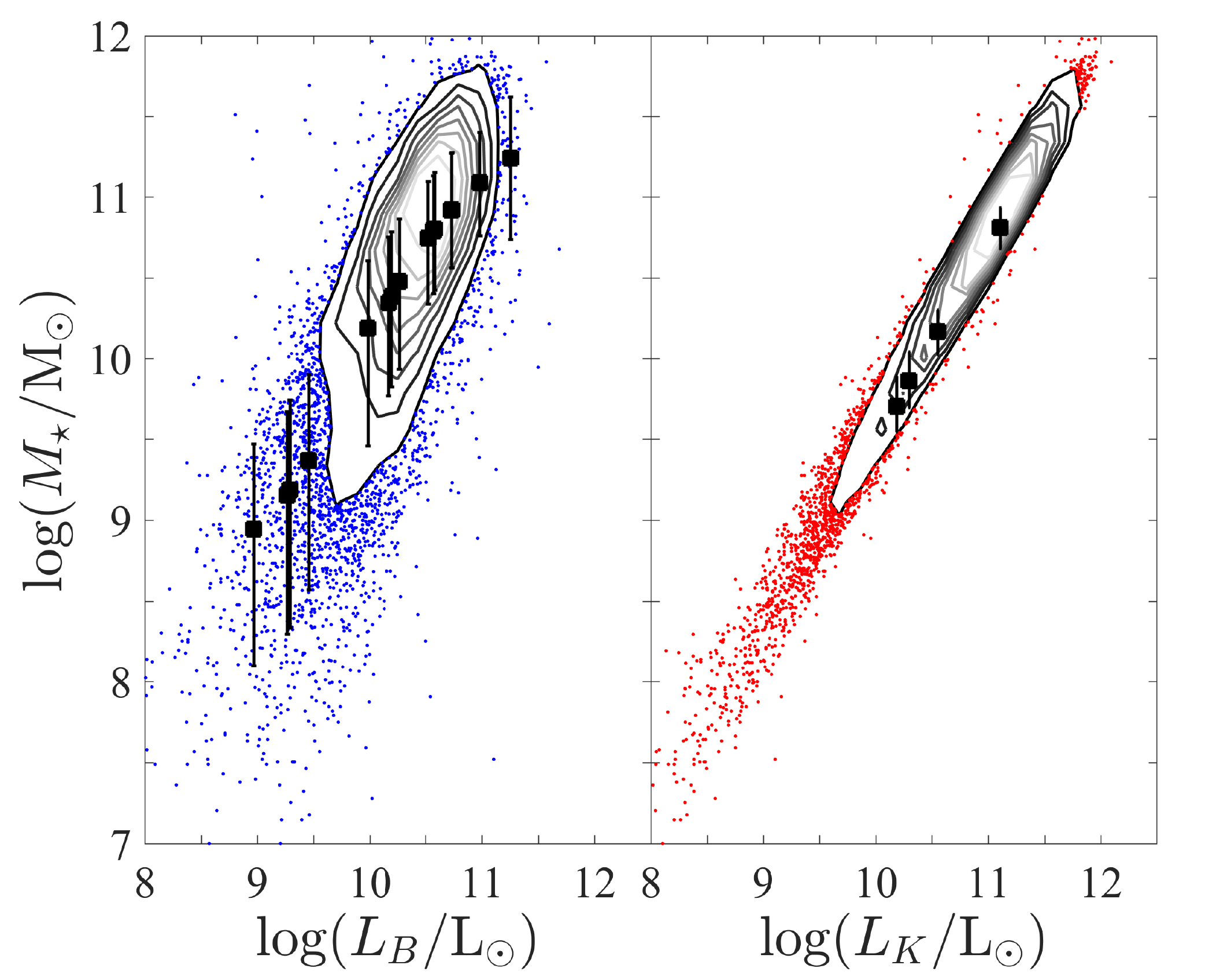}
 \caption{Correlations between the $B$- and $K$-band luminosities of all LOSS SN host galaxies and their stellar masses. The stellar masses are more tightly correlated with the $K$-band, rather than the $B$-band, luminosities. The estimated stellar masses of the 21 LOSS galaxies in the volume-limited sample without such values are shown as black squares. The vertical error bars denote the 16th and 84th percentiles of the distributions of stellar masses in a bin of width 0.2 dex centered on either the $B$- or $K$-band luminosities of those galaxies.}
 \label{fig:mass2}
\end{figure}

\citet{2011MNRAS.412.1419L} used $B$- and $K$-band photometry to estimate the stellar masses, $M_\star$, of the LOSS galaxies (see Appendix A in Paper I), but 21 of the SN host galaxies in the volume-limited sample lacked stellar mass estimates. Discarding the SNe from these galaxies would render the volume-limited sample unusable in this work, as the lack of mass measurements is not correlated with the fractions of the SNe in the sample. Removing any SNe from the sample would render it incomplete, and the fractions of the remaining SNe would no longer correspond to the true fractions in nature. Fortunately, none of the host galaxies lacked both $B$- and $K$-band data. In Figure~\ref{fig:mass2}, we plot the LOSS stellar masses as a function of luminosity, from which it is apparent that the correlation between the stellar masses and the $K$-band luminosity ($L_K$) is tighter than with $L_B$, with Pearson correlation coefficients of 0.99 and 0.75, respectively. 

\begin{figure}
 \includegraphics[width=0.47\textwidth]{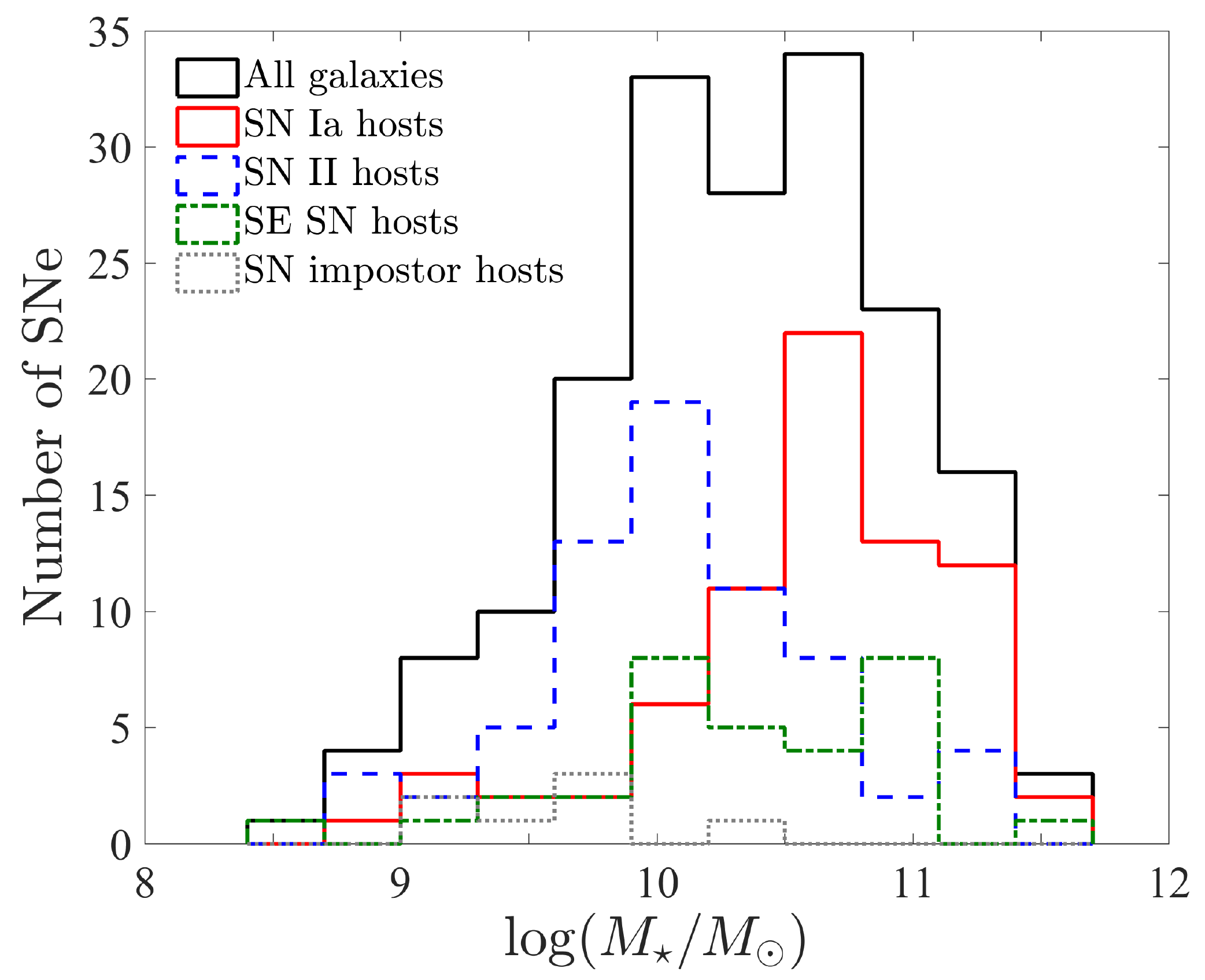}
 \caption{Stellar mass distributions of all SN and SN impostor host galaxies in the volume-limited LOSS sample (black solid). Similar distributions are shown for the hosts of SNe~Ia (red solid), SNe~II (blue dashed), SE~SNe (green dot-dashed), and SN impostors (gray dotted). As in Paper I, all masses have been divided by a factor of $1.2$ to maintain consistency with SDSS Galspec mass values.}
 \label{fig:mass1}
\end{figure}

For each of the 21 galaxies without LOSS masses, we use the median of the distribution of stellar-mass values in a bin of width 0.2 dex centered on either the $L_B$ or $L_K$ luminosity of each galaxy, and take the 16th and 84th percentiles as the uncertainties of the measurement. We note that the galaxies with missing masses are spread throughout the mass range of the sample, so that our estimates of the missing masses should not add a systematic bias to our analysis below. In Figure~\ref{fig:mass1}, we show distributions of the masses for all host galaxies in the volume-limited sample, as well as separately for SNe Ia, SNe II, SE SNe, and SN impostors. 

As in Paper I, all stellar masses measured by LOSS have been divided by $1.2$ to be consistent with masses measured by the MPA-JHU Galspec pipeline\footnote{\url{http://www.sdss3.org/dr9/algorithms/galaxy.php}} \citep{2003MNRAS.341...33K,2004MNRAS.351.1151B,2004ApJ...613..898T} for LOSS galaxies that were also observed by the Sloan Digital Sky Survey \citep{2000AJ....120.1579Y}.


\section{SN Relative Rates}
\label{sec:vol_lim}

The deficiency of SE~SNe in low-mass galaxies observed in Paper~I is also noticeable in the relative rates of different SN types in the volume-limited sample, once the SN host galaxies are split according to their stellar mass, at $10^{10}~{\rm M}_\sun$.

To account for the uncertainty of the mass values that we assigned to the 21 galaxies that did not have such values originally, as well as the uncertainty in the classification of some SNe (e.g., SN 2002ds is either a SN II or a SN IIb), we ran 1000 Monte Carlo simulations to measure the uncertainties of each SN fraction. Each SN was assigned a classification weight: 1 if the classification was certain, or a fraction of 1, depending on the number of possible classifications. To test for the effect of filling in the missing masses, we varied their value according to the probability density function of the distribution of masses from which they were drawn. The SNe then find themselves on either side of the mass threshold.

The final step requires summing the number of SNe of each subtype, dividing by the total number of SNe, and estimating the uncertainties on the resultant fraction. The last step should be a simple ratio of two Poisson distributions, but is complicated by the need to multiply each SN by its individual completeness correction, as measured by L11. To take these corrections into account, we treated each SN as being drawn from a Poisson distribution with an expectation value of $\lambda=1$. Thus, in each step of the Monte Carlo simulation, an individual SN effectively takes on a value drawn from the Poisson distribution. This value is then multiplied by the completeness correction and classification weight of the SN it represents, and summed with the other SNe of its subtype. We took the median of the resulting distribution and divided it by the median of the distribution of all SNe (either SNe Ia or CC SNe). Because of a combination of small-number statistics and of some SNe flipping from one mass bin to the other (if the randomly selected filled-in mass of their host galaxy happens to straddle the mass limit), the median total SNe is sometimes slightly greater than the sum of the medians of the SN distributions. This is only a small difference, however, and does not affect the SN subtypes discussed in detail below.

\floattable
\begin{deluxetable}{lCC|CC|c}
 \tablecaption{Relative SN Rates in the LOSS Sample; Galaxy Stellar Mass Cut $10^{10}~{\rm M}_\sun$. \label{table:fractions}}
 \tablehead{
 \colhead{SN type} & \colhead{Number} & \colhead{Fraction} & \colhead{Number} & \colhead{Fraction} & \colhead{Probability}
 }
 \startdata
               & \multicolumn{2}{c}{$M_\star\leq 10^{10}~{\rm M}_\sun$} & \multicolumn{2}{c}{$M_\star > 10^{10}~{\rm M}_\sun$} & \\
  \multicolumn{6}{c}{All SNe, out to 60 Mpc} \\
  SNe Ia       & 10.0^{+3.5}_{-3.0} & 0.20^{+0.07}_{-0.06} & 27^{+5}_{-5} & 0.27^{+0.06}_{-0.05} & 9.0\% \\ 
  SE SNe       & 9.8^{+3.2}_{-2.7}  & 0.19^{+0.06}_{-0.05} & 28^{+6}_{-5} & 0.29^{+0.06}_{-0.06} & 5.7\% \\
  SNe II       & 31^{+7}_{-6}       & 0.61^{+0.13}_{-0.12} & 42^{+8}_{-7} & 0.44^{+0.08}_{-0.07} & 11\%  \\
  Total        & \multicolumn{2}{c}{50.6}                     & \multicolumn{2}{c}{96.9}                     &       \\
  \multicolumn{6}{c}{CC SNe, out to 60 Mpc} \\
  IIP/L        & 25^{+6}_{-6}       & 0.60^{+0.15}_{-0.14}    & 40^{+7}_{-7}       & 0.6^{+0.1}_{-0.1}       & 18\%  \\ 
  II-87A       & 3.0^{+2.0}_{-2.0}  & 0.08^{+0.05}_{-0.05}    & 0.0^{+1.8}_{-0.0}  & 0.000^{+0.025}_{-0.000} & 1.4\% \\
  IIn          & 3.0^{+2.0}_{-2.0}  & 0.07^{+0.05}_{-0.05}    & 2.0^{+1.0}_{-1.0}  & 0.028^{+0.014}_{-0.014} & 10\%  \\
  IIb          & 4.6^{+2.3}_{-2.0}  & 0.11^{+0.06}_{-0.05}    & 7.6^{+2.8}_{-2.6}  & 0.11^{+0.04}_{-0.04}    & 30\%  \\
  Ib           & 0.9^{+0.7}_{-0.5}  & 0.022^{+0.017}_{-0.012} & 10.1^{+4.1}_{-3.5} & 0.14^{+0.06}_{-0.05}    & 1.3\% \\
  Ic           & 1.8^{+1.3}_{-1.1}  & 0.044^{+0.032}_{-0.027} & 5.0^{+2.0}_{-2.0}  & 0.07^{+0.03}_{-0.03}    & 25\%  \\ 
  (Ib$+$Ic)\tablenotemark{a}      & 2.7^{+1.5}_{-1.2}  & 0.07^{+0.04}_{-0.03}      & 15^{+5}_{-4}            & 0.22^{+0.07}_{-0.06} & 1.4\% \\
  IIb-pec      & 0.6^{+0.6}_{-0.6}  & 0.015^{+0.015}_{-0.015} & 0.0^{+1.8}_{-0.0}  & 0.000^{+0.025}_{-0.000} & 34\%  \\
  Ic-pec       & 0.0^{+1.8}_{-0.0}  & 0.00^{+0.04}_{-0.00}    & 1.0^{+1.0}_{-1.0}  & 0.014^{+0.014}_{-0.014} & 34\% \\
  Ic-BL        & 0.4^{+0.4}_{-0.4}  & 0.01^{+0.01}_{-0.01}    & 1.5^{+1.0}_{-1.0}  & 0.021^{+0.014}_{-0.014} & 61\%  \\
  Ca-rich      & 1.3^{+1.3}_{-1.3}  & 0.03^{+0.03}_{-0.03}    & 2.4^{+1.4}_{-1.3}  & 0.034^{+0.020}_{-0.018} & 41\%  \\
  Total        & \multicolumn{2}{c}{40.8}                     & \multicolumn{2}{c}{70.6}                     &       \\
  \multicolumn{6}{c}{SNe Ia, out to 80 Mpc} \\
  IaN          & 10^{+3}_{-3}       & 0.87^{+0.26}_{-0.26}    & 43^{+7}_{-7}       & 0.7^{+0.1}_{-0.1}       & 13\%  \\
  Ia-99aa      & 0.5^{+0.5}_{-0.5}  & 0.04^{+0.04}_{-0.04}    & 4.5^{+2.0}_{-2.0}  & 0.07^{+0.03}_{-0.03}    & 62\%  \\
  Ia-91T       & 0.0^{+1.8}_{-0.0}  & 0.00^{+0.16}_{-0.00}    & 1.0^{+1.0}_{-1.0}  & 0.016^{+0.016}_{-0.016} & 61\%  \\
  Ia-91bg      & 0.0^{+1.8}_{-0.0}  & 0.00^{+0.16}_{-0.00}    & 11^{+3}_{-3}       & 0.18^{+0.05}_{-0.05}    & 6.7\% \\
  Ia-02cx      & 1.0^{+1.0}_{-1.0}  & 0.09^{+0.09}_{-0.09}    & 1.0^{+1.0}_{-1.0}  & 0.016^{+0.016}_{-0.016} & 13\%  \\
  Ia-02es      & 0.0^{+1.0}_{-0.0}  & 0.00^{+0.09}_{-0.00}    & 1.0^{+2.0}_{-1.0}  & 0.016^{+0.032}_{-0.016} & 61\%  \\
  Total        & \multicolumn{2}{c}{11.5}                     & \multicolumn{2}{c}{62.7}                     &       \\
 \enddata   
 \tablenotetext{a}{Because the classification of some SNe as either SNe Ib or Ic is uncertain, we also present the fraction of both types, combined.}
\end{deluxetable}

\begin{figure*}
 \center
 \plotone{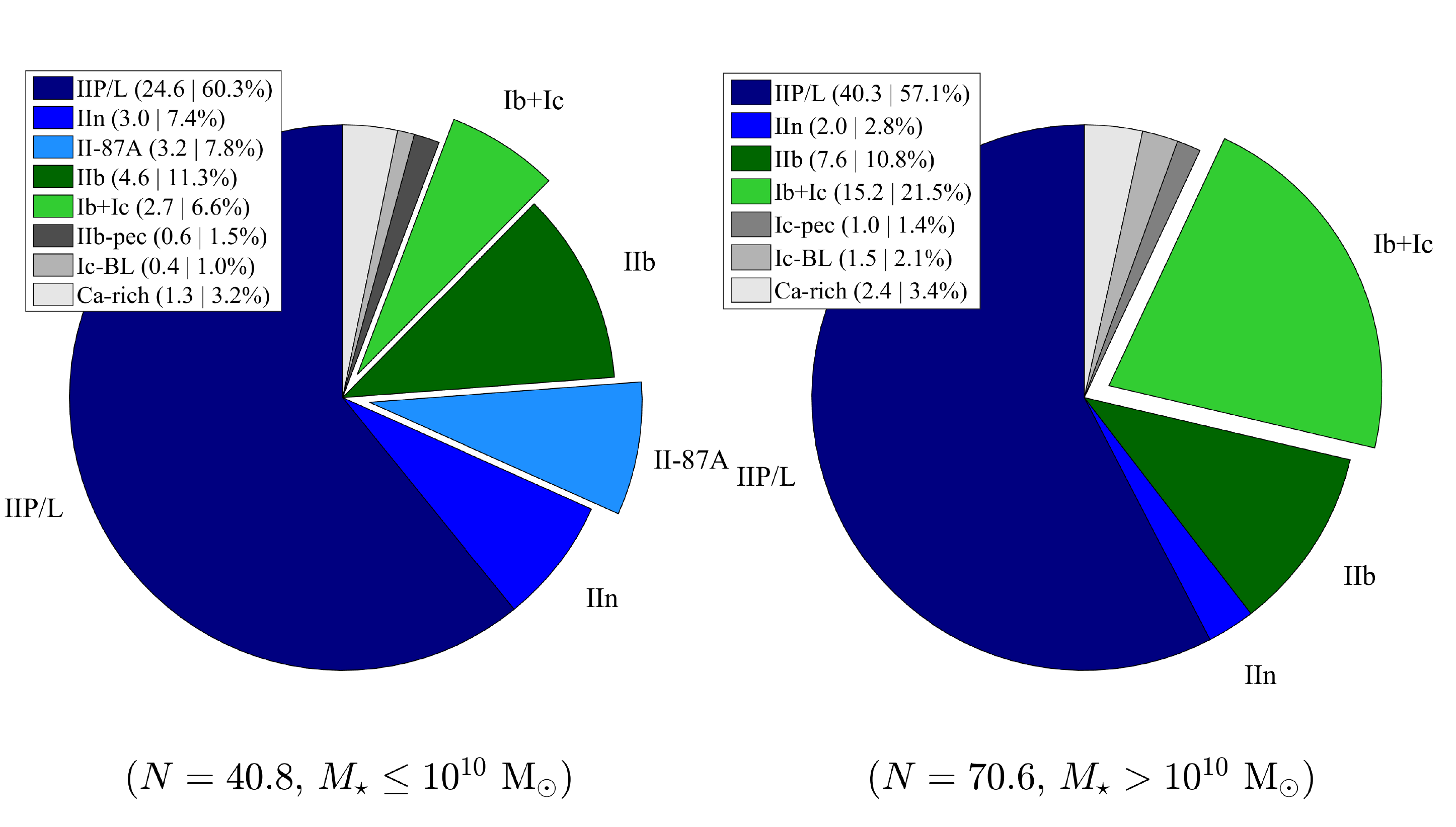} \\
 \plotone{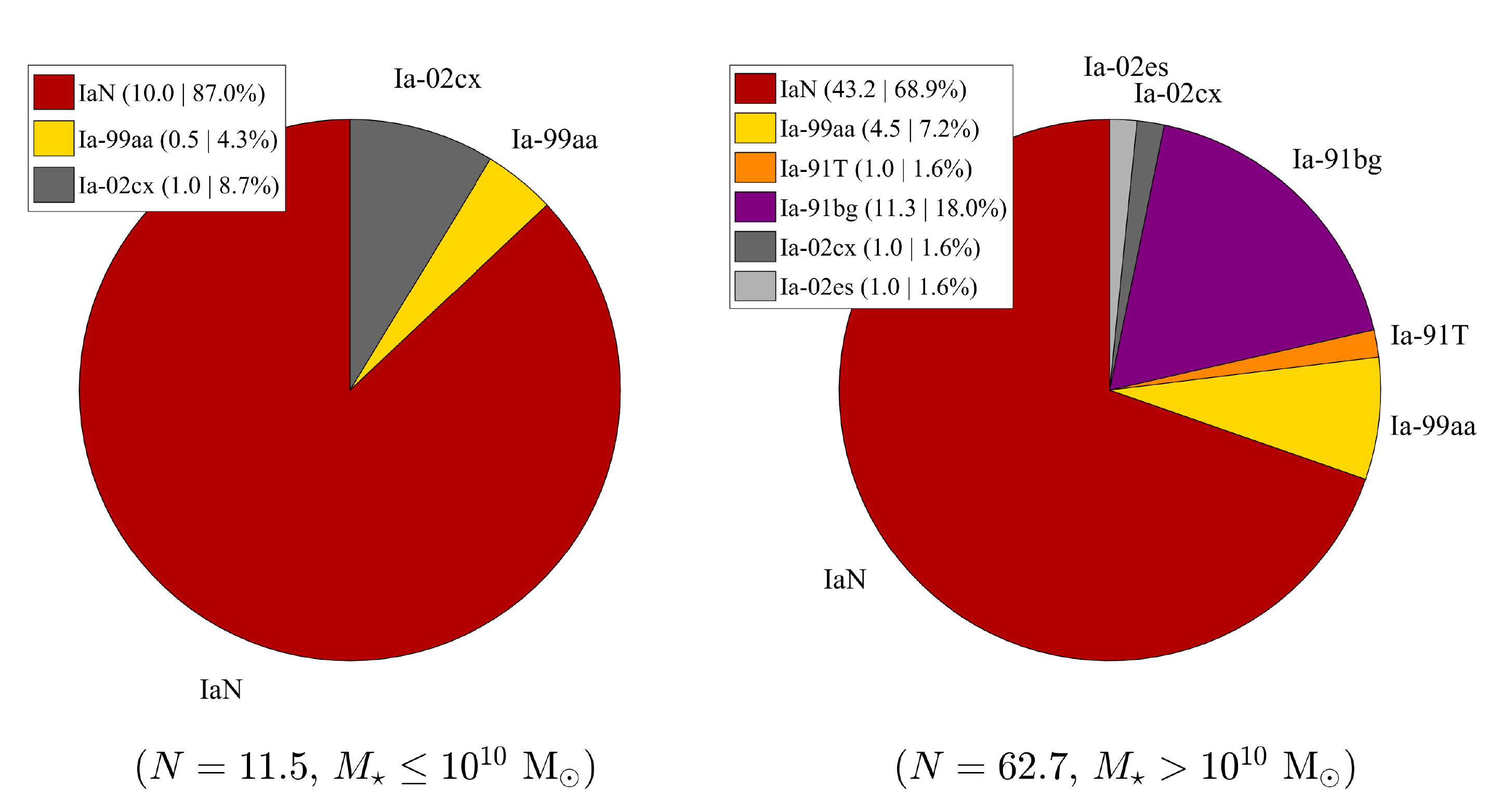}
 \caption{Relative SN rates in the LOSS volume-limited sample of CC SNe, complete out to 60 Mpc (top), and SNe Ia, complete to 80 Mpc (bottom). The left and right pie charts show the fractions of SNe in galaxies with stellar masses $\leq10^{10}~{\rm M}_\sun$ and $>10^{10}~{\rm M}_\sun$, respectively. This is roughly the stellar mass at which the absolute SE SN rates begin to deviate appreciably from the SN II rates, as shown in Paper I. The legends denote the number of SNe of each subtype, after correcting for completeness, classification, and stellar mass uncertainties (thus giving rise to the non-integer numbers of SNe of different types), and the resulting fractions, as percentages. The fractions here represent the medians of the distributions generated by Monte Carlo simulations for each SN subtype. Unlabeled slices represent peculiar SNe, and are listed in the legends. Exploded slices show SN subtypes for which the fractions in each of the galaxy samples differ in a statistically significant manner ($5\%$ significance). Namely, there are three SN 1987A-like SNe in low-mass galaxies, but none in the more massive galaxies; and there are roughly three times fewer SNe Ib and Ic, combined, in the less massive galaxies than in the more massive ones. Although not statistically significant, there are roughly $1.3$ times as many normal SNe Ia in low-mass galaxies. Subluminous SN 1991bg-like SNe~Ia, which only appear in massive galaxies, seem to cut into the fraction of normal SNe~Ia, which otherwise would be nearly identical to the fraction of normal SNe~Ia in less massive galaxies. See Table~\ref{table:fractions} for 68\% uncertainties on the fractions shown here.}
 \label{fig:pie}
\end{figure*}

The resulting fractions of the different SN types in the LOSS volume-limited sample are shown in Figure~\ref{fig:pie} and presented in Table~\ref{table:fractions}. The fourth column of Table~\ref{table:fractions} lists the probability of randomly discovering the recorded number of SNe of each type, assuming that all galaxies host the same intrinsic fractions of SN types. As in \citet{2010ApJ...721..777A}, the probability of measuring an underabundance, $n$, of SNe of a given subtype in a sample of $N$ galaxies (or an overabundance, $l$, of SNe of the same subtype in the other $L$ galaxies) is calculated according to
\begin{equation}\label{eq:likelihood}
 \sum_{x=0}^{n} \binom{N}{x} p^x (1-p)^{N-x} \sum_{y=l}^{L} \binom{L}{y} p^y (1-p)^{L-y},
\end{equation}
where we vary the intrinsic SN fraction ($p$) between 0 and 1, assuming that it is independent of the host galaxy's stellar mass. A difference between relative rates of a given SN subtype (and hence, a rejection of the assumption that $p$ is the same in all galaxy types) is considered statistically significant if the maximal probability of it occuring by chance is $\le 5\%$.

Based on the resulting probabilities, we find that as one family, SNe II are overrepresented in low-mass galaxies. SE SNe and SNe Ia may be underrepresented in these galaxies, but the samples are too small for a definitive conclusion.

Examining the CC SN fractions in detail, we find that low-mass galaxies host the only three SN 1987A-like objects in the sample (consistent with \citealt{2016A&A...588A...5T}, who find that this type of SN is preferentially found in low-metallicity galaxies).

SNe Ib are underrepresented, by a factor of $\sim 5.7$, in galaxies with $M_\star \leq 10^{10}~{\rm M}_\sun$. The fraction of SNe Ic in low-mass galaxies is formally smaller by a factor of $\sim 1.5$, although the difference is not statistically significant. As classifying SNe as either SNe Ib or Ic is not always straightforward (e.g., \citealt{2014AJ....147...99M,2016ApJ...827...90L,2016arXiv160902922S}), we also calculate the fractions of SNe Ib and Ic, combined. The latter fraction is $\sim 2.9$ times smaller in low-mass galaxies and is statistically significant. This result is consistent with our finding in Paper I that the ratio between the SE SN and SN II absolute rates is lower in low-mass galaxies by a factor of $\sim3$ relative to the ratio in high-mass galaxies. 

\citet{2016arXiv160902922S} found that the ratio of SNe Ib to SNe Ic is $1.5\pm0.7$. From Table~\ref{table:fractions}, it would seem that this ratio, which is $\sim2$ in high-mass galaxies, is reversed in low-mass galaxies. We caution not to attribute any significance to this reversal, as the low-mass galaxies contain only one SN Ib and two SNe Ic, leading to a ratio of $0.5^{+0.5}_{-0.5}$. This ratio is consistent, at $2\sigma$, with the SN Ib/Ic ratio of $2.0^{+2.0}_{-0.8}$ in high-mass galaxies.

Intriguingly, the SN IIb fractions are consistent between the two galaxy populations. We note that it should be as easy to discover SNe IIb as it is to discover SNe Ib and Ic, as their luminosity functions and light curves are similar (e.g., L11; \citealt{2011ApJ...741...97D,2012ApJ...756L..30A,2016MNRAS.457..328L}; but see \citealt{2015A&A...574A..60T} regarding faster rise times for SNe Ic and Ic-BL, relative to SNe Ib and IIb). We therefore do not think that the different trends for SNe Ib+Ic and SNe IIb could be due to LOSS having systematically missed any of these SN subtypes.

L11 did not compute completeness corrections for the SN impostors in the volume-limited sample, therefore they are not included in Figure~\ref{fig:pie} and Table~\ref{table:fractions}. It is still interesting to note, however, that of the seven SN impostors classified as such by L11 and \citet{2011MNRAS.415..773S}, six are in galaxies with $M_\star<10^{10}~{\rm M}_\sun$.\footnote{The host galaxy of SN 2002bu does not have a LOSS stellar mass value. As explained in Section~\ref{sec:galaxies}, we esitmate a mass of $\sim 4\times10^9~{\rm M}_\sun$.} If we were to count these objects as CC SNe (as they could be CC SN progenitors at the end of their lives), then, without completeness corrections, they would account for $12.3$\% and $1.4$\% of CC SNe in low- and high-mass galaxies, respectively. In other words, SN impostors are overrepresented in low-mass galaxies by a factor of $\sim 9$, a statistically significant result (with a probability of $0.5$\%). This result is consistent with \citet{2015A&A...580A.131T}, who find that SN impostors explode in lower-metallicity environments than SNe~IIn. However, in our sample this result can also be due to a selection effect, as SN impostors are generally fainter than SNe and so would be easier to pick out in lower-mass, lower-luminosity galaxies.

Although not statistically significant, normal SNe~Ia are overrepresented in low-mass galaxies by a factor of $\sim 1.3$. We note that while the lack of SN 1991bg-like SNe~Ia in the low-mass galaxies is consistent with the large fraction (18\%) found in high-mass galaxies, in the latter, the SN 1991bg-like SNe~Ia seem to cut into the fraction of normal SNe~Ia. It will be interesting to see whether this effect persists in a larger sample. If it does, it reinforces the suggestion made by \citet{2015MNRAS.450..905G} that in order to collect a homogeneous sample of SNe Ia for cosmological purposes, one would be best served targeting low-mass galaxies, as opposed to star-forming galaxies---without regard to their mass---as advocated by \citealt{2014MNRAS.445.1898C}. From an observational point of view, this would best be achieved by targeting low-luminosity (i.e., low-mass) galaxies.

To test the sensitivity of our results to the value of the mass cut, we repeated the above exercise with two different mass cuts: $3\times10^9~{\rm M}_\sun$ and $3\times10^{10}~{\rm M}_\sun$ (the first value is motivated by the mass of the Large Magellanic Cloud, or LMC, which is often treated as the upper mass limit of dwarf galaxies; \citealt{2002AJ....124.2639V}). The resulting relative rates are presented in Tables~\ref{table:fractions_mass3e9} and \ref{table:fractions_mass3} and Figures~\ref{fig:pie_3e9} and \ref{fig:pie_3e10}.

When using the higher mass cut at $3\times10^{10}~{\rm M}_\sun$, we find that SNe Ia are significantly underrepresented in low-mass galaxies by a factor of $\sim 2.4$ and SNe~II are overrepresented by a factor of $\sim 2.2$, whereas before they were consistent between the two samples. SNe Ib and Ic (individually and combined) are still deficient in low-mass galaxies by a factor of $\sim3$ (this result also becomes significant for SNe Ic). Although the formal ratio between SNe IIb in high- and low-mass galaxies is $\sim 1.5$, the two relative rates are still consistent (whereas in \citealt{2010ApJ...721..777A} they were underrepresented by a factor of 5, as we discuss in Section~\ref{subsec:vol_PTF}, below). SN 1987A-like SNe are still only found in low-mass galaxies, but this result is no longer significant. The difference between normal SNe Ia and SN 1991bg-like SNe~Ia in low- and high-mass galaxies becomes statistically significant.

The lower mass cut at $3\times10^9~{\rm M}_\sun$ reduces the number of SNe in low-mass galaxies to the point where only one difference between the galaxy samples is significant: an overabundance of SNe IIn in low-mass galaxies by a factor of $\sim 6$. Formally, SNe~IIb are roughly twice as common in low-mass galaxies, but the relative rates of this SN subtype are still consistent between the two galaxy samples.

The three mass cuts used here show that the most robust trend is the deficiency of SNe Ib and Ic in low-mass galaxies. This trend is also qualitatively present when the lowest mass cut is used, but is no longer significant as the low-mass SN sample is reduced to the point where it is dominated by Poisson noise. The following trends also stand out regardless of the mass cut used, but are only significant when one of the mass cuts is used: (1) normal SNe Ia are overabundant in low-mass galaxies; (2) SN 1991bg-like SNe~Ia are only found in high-mass galaxies; and (3) SN 1987A-like SNe prefer low-mass galaxies. Finally, the relative rates of SNe IIb are consistent with each other in both low- and high-mass galaxies, independent of which mass cut is used.

\subsection{Caveats}
\label{subsec:vol_caveats}

As noted in Section~\ref{sec:galaxies}, the targeted nature of LOSS means that the volume-limited sample, while complete to all SN types out to 60 Mpc, is incomplete to all galaxy types. We do not think that this incompleteness is the source of the deficiency we observe in the SN Ib and SN Ic relative rates in low-mass (and hence low-luminosity) galaxies. First, as \citet{2011MNRAS.412.1419L} note, low-luminosity galaxies, which have low star-formation rates, do not produce many SNe. They estimate that only $\sim 15$\% of CC SNe and $\sim 10$\% of SNe Ia are missed in the volume-limited sample. Second, as Figure~\ref{fig:mass1} shows, the LOSS volume-limited sample is not bereft of low-mass galaxies: $28.3$\% ($28.9$\% once the missing masses are filled in) of the host galaxies have stellar masses $<10^{10}~{\rm M}_\sun$. Third, we have no reason to think that the low-luminosity galaxies included in LOSS, and the SNe they hosted, are in any way uncharacteristic of the overall low-luminosity galaxy population, which means that the rates we measure, while suffering from large statistical uncertainties, should not be systematically biased by the galaxy sample's incompleteness. 

We used the L11 completeness corrections, which were calculated using light-curve templates (also constructed by L11), based on the classification of each SN. Although some of the classifications have changed, we did not recalculate the corrections, as (1) only a small fraction of the classifications have changed, and (2) the luminosity functions and light curves of SNe Ib, Ic, and IIb (the SN types that were most affected by the classification changes in \citealt{2016arXiv160902922S}) are very similar.

\citet{2013ApJ...767...57F} have argued that the relative rates of SNe Iax measured by L11 were underestimated, as the SNe Iax in the LOSS sample were all high-luminosity objects (brighter than $-16.7$ mag at peak). Since we rely on the L11 completeness corrections, our fractions of SNe Iax suffer from the same bias. However, as in this paper we compare relative SN rates in galaxies of different masses, and as the differences we find for SNe Iax are not statistically significant, we do not attempt to correct for this bias. 

\subsection{Comparison with the Palomar Transient Factory}
\label{subsec:vol_PTF}

Our results are in some tension with those of the PTF. Below, we compare our results to the preliminary rates presented by \citet{2010ApJ...721..777A} and to the updated rates shown by \citet{2012IAUS..279...34A}. As we discuss below, it is hard to evaluate the significance of the updated results from \citet{2012IAUS..279...34A}, as they are only shown---not discussed---in the conference proceedings. If the latter results hold up, then the LOSS rates differ from those of the PTF only when it comes to SNe~IIb: PTF find that these SNe are more abundant in dwarf galaxies, while we find no significant differences between the SN~IIb population in low-mass and high-mass galaxies.

\citet{2010ApJ...721..777A} split a preliminary sample of 70 PTF SNe among ``giant'' and ``dwarf'' host galaxies, defined by the galaxies' $r$-band luminosities. Our findings contradict theirs in three main points, as follows.
\begin{enumerate}
 \item They found that the overall fraction of all SE SN types is larger in low-mass galaxies; we find the opposite.
 \item They found no difference between the relative rates of SNe Ib in giants and dwarfs, while we find a marked deficiency of SNe Ib in low-mass galaxies.
 \item They found a deficiency of SNe IIb in giant galaxies. We find no difference between the fractions of this SN subtype in low- and high-mass galaxies. 
\end{enumerate}

Additionally, \citet{2010ApJ...721..777A} found SNe Ic-BL to be overrepresented in dwarf galaxies, where they accounted for 13\% of all SNe. This result was based on two SNe Ic-BL in dwarf galaxies and one in giant galaxies. Our sample includes a similar number of 1--2 SNe Ic-BL, but we find no significant difference between the fractions of this SN in the two galaxy types. It is interesting to note that as a result of the makeup of our SN sample, SNe Ic-BL only account for $\sim1\%$ of all SNe in low-mass galaxies, an order of magnitude lower than in \citet{2010ApJ...721..777A}. However, owing to small-number statistics, we caution that this difference between the two studies may not be significant.

Although the LOSS volume-limited sample is slightly larger than the PTF sample (99 CC SNe vs. 70), any conclusions drawn from the LOSS and PTF samples are both constrained by small-number statistics. However, the LOSS volume-limited sample is better suited for this prupose, as it represents a complete sample and includes completeness corrections based on the survey's sensitivity to each SN in the sample. The PTF sample, on the other hand, is incomplete, as its detection efficiency for different types of SNe is unknown. It is also unknown how well the PTF SN host galaxies represent the local galaxy luminosity function. That said, whereas the LOSS survey targeted massive galaxies, the PTF survey was untargeted and therefore more sensitive to dwarf galaxies. 

Another difference between the PTF and LOSS samples is their classification completeness, i.e., how many spectra (and at what phases) are used to classify the SNe. This is most important for the classification of SNe IIb. Many of the spectroscopic differences between SNe Ib and IIb may be time-dependent (e.g., \citealt{1988AJ.....96.1941F,1993ApJ...415L.103F,1997ARA&A..35..309F,2008MNRAS.389..113P,2011ApJ...739...41C,2013ApJ...767...71M}). PTF obtained multiple spectra for SNe that were initially classified as SNe II to check if they later evolved into SNe IIb. They also checked early spectra of SNe Ib for hydrogen features (I. Arcavi, private communication).

Unfortunately, many of the events within the LOSS sample have only partial spectroscopic coverage. However, \citet{2016ApJ...827...90L} show that the population of SNe IIb exhibit characteristically stronger hydrogen lines than the SNe Ib population at all photospheric phases, and (using the updated classification tools of \citealt{2014arXiv1405.1437L}) \citet{2016arXiv160902922S} were able to distinguish between these two subtypes in most examples. For only one SN in the LOSS volume-limited sample is a clear distinction between Ib and IIb impossible, and it is classified as Ib/IIb. The time-variance of our ability to spectroscopically distinguish between type II and IIb SNe, however, is less well understood, and because of the LOSS sample's only partial spectroscopic coverage, there are five SNe in the sample that may be of type II or IIb. These classification uncertainties have been taken into account here, as detailed in Section~\ref{sec:vol_lim}.

Finally, it is important to note that in this work we split the SN host galaxies by their stellar mass, and define low-mass galaxies as having stellar masses $<10^{10}~{\rm M}_\sun$. This threshold includes both ``dwarf'' galaxies, which are often defined as being less massive than the LMC at $3\times10^9~{\rm M}_\sun$, and some ``giant'' galaxies. \citet{2010ApJ...721..777A} split their sample according to the $r$-band luminosity of the galaxies, with dwarf galaxies defined as being fainter than $M_r \ge -18$~mag. Of the galaxies in the LOSS volume-limited sample, $10.1$\% ($11.1$\% when missing masses are filled in) can be defined as dwarfs according to their mass. On the other hand, $21.4$\% of the PTF galaxies are defined as dwarfs. This shows that while the LOSS sample does probe SNe in dwarf galaxies, the PTF sample is more sensitive to them.

An update of the preliminary PTF results appeared in figure 3 of \citet{2012IAUS..279...34A}, where the SN sample had been enlarged to 369 objects. The larger sample led to some changes in the preliminary trends, as follows:
\begin{enumerate}
 \item SNe Ib and SNe Ic might be overrepresented in more luminous (and hence more massive) galaxies, as the relative rates for the most luminous galaxies are twice and three times, respectively, those of the rates in the least luminous ones. This is consistent with the rates we measure here.
 \item The SN IIb rate is still lower in more luminous massive galaxies, than in less luminous low-mass galaxies. This is still in contrast with our finding that SNe IIb seem to have no preference for either low- or high-mass galaxies.
 \item There is no apparent trend between the fraction of SNe Ic-BL, relative to other CC SNe, and the luminosity of their host galaxies. This is consistent with our findings.
\end{enumerate}

Although the updated PTF results are based on a larger sample of SNe, there are some caveats to keep in mind. First, \citet{2012IAUS..279...34A} does not describe the composition of the SN sample (e.g., classification methods) or its completeness. Second, there is no quantitative discussion of the measurements shown in figure 3. While it is possible to claim ``by eye'' the above trends for SNe IIb, Ib, and Ic, the measurements seem to be---again, ``by eye''---consistent with each other at the $2\sigma$ level. Thus we cannot tell if there are any significant trends in the updated measurements; there could just as easily be no trends at all.

At the time of writing, we have been informed that an in-depth analysis of the complete PTF SN sample---including new rate measurements---is under way (A. Gal-Yam, private communication). It will be interesting to see whether the final results from the full PTF sample, as well as future rates from the All-Sky Automated Survey for Supernovae (ASAS-SN; \citealt{2017MNRAS.464.2672H}) will confirm or contradict the LOSS rates described here.


\section{Summary and Conclusions}
\label{sec:discuss}

This is the second of a series of papers that further explore the implications of the LOSS SN rates. Here, we examined the relative rates of different SN subtypes in the volume-limited LOSS sample. This sample, which contains 180 SNe and SN impostors, is complete for SNe Ia out to 80 Mpc and CC SNe out to 60 Mpc. Our analysis was based on a reclassification of the SNe in this sample \citep{2016arXiv160902922S}. Where L11 originally did not distinguish between galaxy types, we split the LOSS volume-limited sample into two galaxy samples, using different mass cuts, and compared the relative rates of different SN subtypes in low- and high-mass galaxies. This split was motivated by our finding in Paper I that the absolute rates of SE SNe, relative to SNe II, are lower in low-mass galaxies by a factor of $\sim3$. We have found the following trends (Figure~\ref{fig:pie}).
\begin{enumerate}
 \item SNe Ib are underrepresented in low-mass galaxies by a factor of 3--6. Taken together, SNe Ib and Ic are underrepresented in low-mass galaxies by a factor of $\sim 3$. Both results are statistically significant, given our significance threshold of 5\%.
 \item On the other hand, the relative rates of SNe IIb are consistent with each other in both low- and high-mass galaxies.
 \item SN 1987A-like SNe prefer low-mass galaxies.
 \item SN impostors, many of which could signify the death throes of CC SN progenitors, are overrepresented in low-mass galaxies by a factor of $\sim 9$. This could be due to selection effects, however, as the SN impostors in the LOSS volume-limited sample have not been corrected for completeness, and these objects tend to have lower peak luminosities than SNe.
 \item Normal SNe Ia are overrepresented in low-mass galaxies by a factor of $\sim 1.3$. 
 \item As expected, subluminous SN 1991bg-like SNe~Ia are only found in high-mass galaxies, but intriguingly, they cut into the share of the normal SNe Ia. 
\end{enumerate}
We find that the first trend is the most robust to the choice of mass cut, followed by trends 3, 5, and 6. Although not formally statistically significant, trend 2 is qualitatively present regardless of which mass cut is used. 

The underrepresentation of SNe Ib and Ic in low-mass galaxies strengthens our finding in Paper I that the SE SN rates, relative to the SN II rates, are lower in low-mass galaxies than in high-mass galaxies. In the latter type of galaxy, we find the same fraction of SNe Ib and Ic, combined, as \citet{2011MNRAS.412.1522S} did. This strengthens their point that single stars, on their own, cannot account for the observed fractions of SE SNe, at least in high-mass galaxies. This is true if one were to assume that all galaxies have the same IMF. Recent studies, however, have reported that galaxies with higher star-formation rates show more top-heavy IMFs (e.g., \citealt{2011MNRAS.415.1647G,2013ApJ...771...29G,2013MNRAS.436.3309W,2014MNRAS.442.1003P}). We leave it to a future paper to explore whether the larger fraction of massive stars produced by such IMFs could account for the higher fraction of SNe Ib and Ic we observe in high-mass (and hence highly star-forming) galaxies. 

It is interesting that the relative rates of SNe IIb are consistent between low- and high-mass galaxies. This adds another reason why the exclusion of this type of SN from the SE SN rates in Paper I should not bias our findings in that paper. It also raises the question whether SNe IIb come from different progenitors than other SE SNe. If, instead, SNe IIb, Ib, and Ic all come from the same type of progenitor stellar system, it remains to be seen what property (or combination of properties, such as metallicity, IMF, or binarity fraction) of galaxies at $M_\star \lesssim 10^{10}~{\rm M}_\sun$ could hinder the production of SNe Ib and Ic, but not of SNe IIb. Our result is in tension with binary evolution models, which predict that the SN~IIb rate should indeed be higher in low-mass 
low-metallicity galaxies---as claimed by the PTF---since line-driven winds should be less efficient at removing the residual H envelope of the low-metallicity SN~IIb progenitor \citep{2011A&A...528A.131C,2014ARA&A..52..487S}. A larger sample of SE~SNe in low-mass galaxies is required to test whether at a lower mass cut the fractions of SNe~IIb will remain the same, will begin to exhibit the type of deficiency we have shown here for SNe~Ib and Ic, or become overabundant, as claimed by the PTF.

The overrepresentation of normal SNe~Ia in low-mass galaxies strengthens the conclusion of \citet{2015MNRAS.450..905G} that in order to construct homogeneous SN Ia samples for cosmology, it would be best to select them from low-mass (or low-luminosity) galaxies, and not just from star-forming galaxies, as suggested by \citet{2014MNRAS.445.1898C}.

The results of this work differ from those of \citet{2010ApJ...721..777A} and \citet{2012IAUS..279...34A}, especially in regard to SNe~IIb. Our samples differ as well, as LOSS is a targeted survey, while PTF is untargeted. We argue here that the LOSS volume-limited sample, with its known sensitivity and completion caveats, is better suited for studying relative rates than the preliminary PTF sample used by \citet{2010ApJ...721..777A}. Still, as the samples from both studies suffer from small-number statistics, we look forward to our results being tested by ongoing and future untargeted surveys.


\section*{Acknowledgments}

We thank the anonymous referee for helpful suggestions and comments. We also thank Iair Arcavi, Ryan Chornock, Avishay Gal-Yam, Jenny Greene, Patrick Kelly, Dan Maoz, Asaf Pe'er, Michael Shara, and Todd Thompson for helpful discussions and comments. 

O.G. is supported in part by National Science Foundation (NSF) award AST-1413260 and by an NSF Astronomy and Astrophysics Fellowship under award AST-1602595. F.B.B. is supported in part by the NYU/CCPP James Arthur Postdoctoral Fellowship. M.M. is supported in part by NSF CAREER award AST-1352405 and by NSF award AST-1413260. A.V.F. and the Lick Observatory Supernova Search (LOSS) have received generous financial assistance from the TABASGO Foundation, US Department of Energy (DoE) SciDAC grant DE-FC02-06ER41453, DoE grant DE-FG02-08ER41653, and many NSF grants (most recently AST-1211916). KAIT and its ongoing operation were made possible by donations from Sun Microsystems, Inc., the Hewlett-Packard Company, the AutoScope Corporation, the Lick Observatory, the NSF, the University of California, the Sylvia \& Jim Katzman Foundation, and the TABASGO Foundation. Research at Lick Observatory is partially supported by a generous gift from Google. We thank the Lick staff for their assistance at the observatory.

This research has made use of NASA's Astrophysics Data System and the NASA/IPAC Extragalactic Database (NED) which is operated by the Jet Propulsion Laboratory, California Institute of Technology, under contract with NASA. We also acknowledge the usage of the HyperLeda database (\url{http://leda.univ-lyon1.fr}).


\software{MATLAB, MPA-JHU Galspec pipeline \citep{2003MNRAS.341...33K,2004MNRAS.351.1151B,2004ApJ...613..898T}.}



\appendix

\section{The effect of different mass cuts}
\label{appendix:cuts}

\begin{figure*}[b!]
 \center
 \plotone{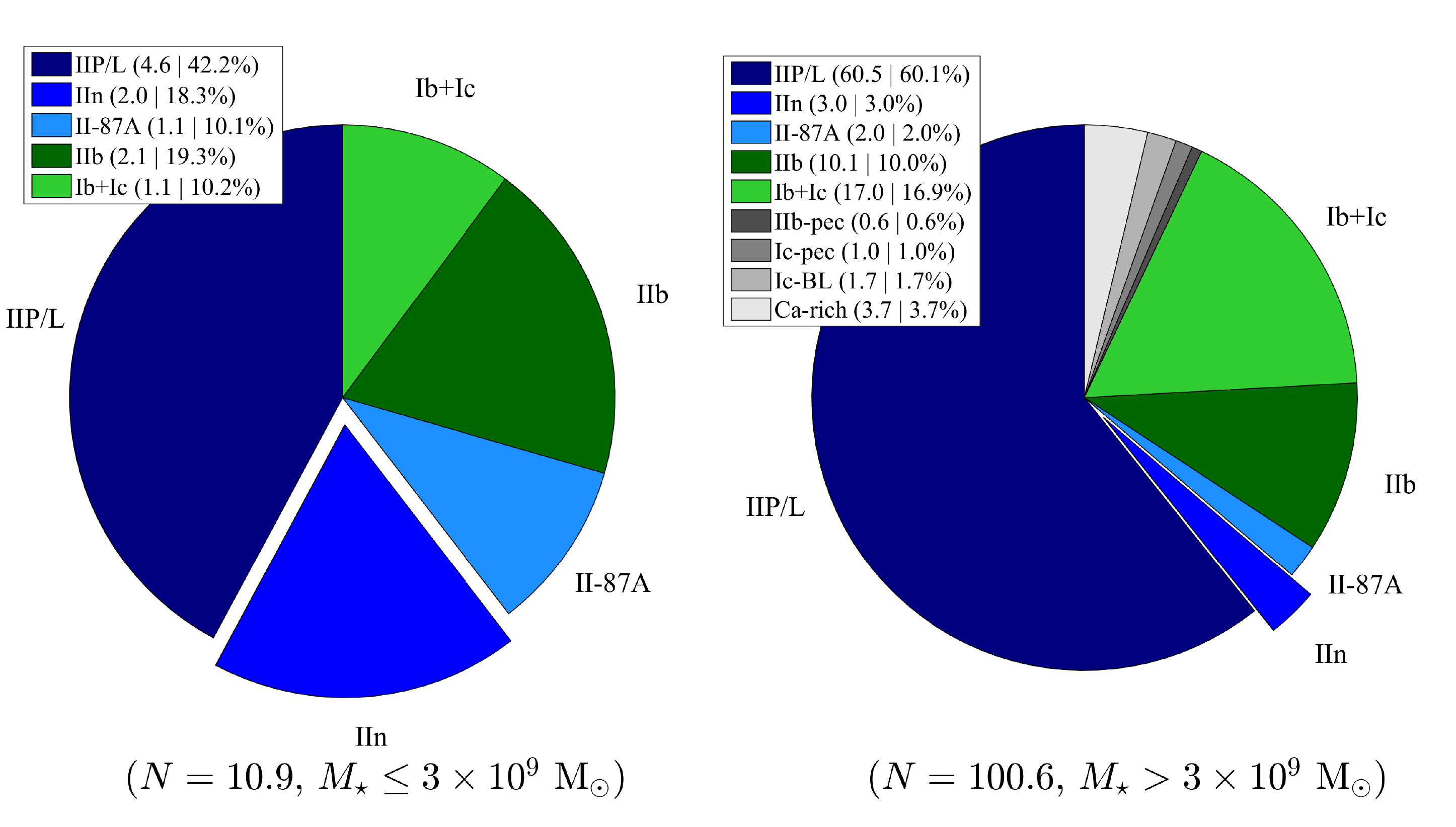} \\
 \plotone{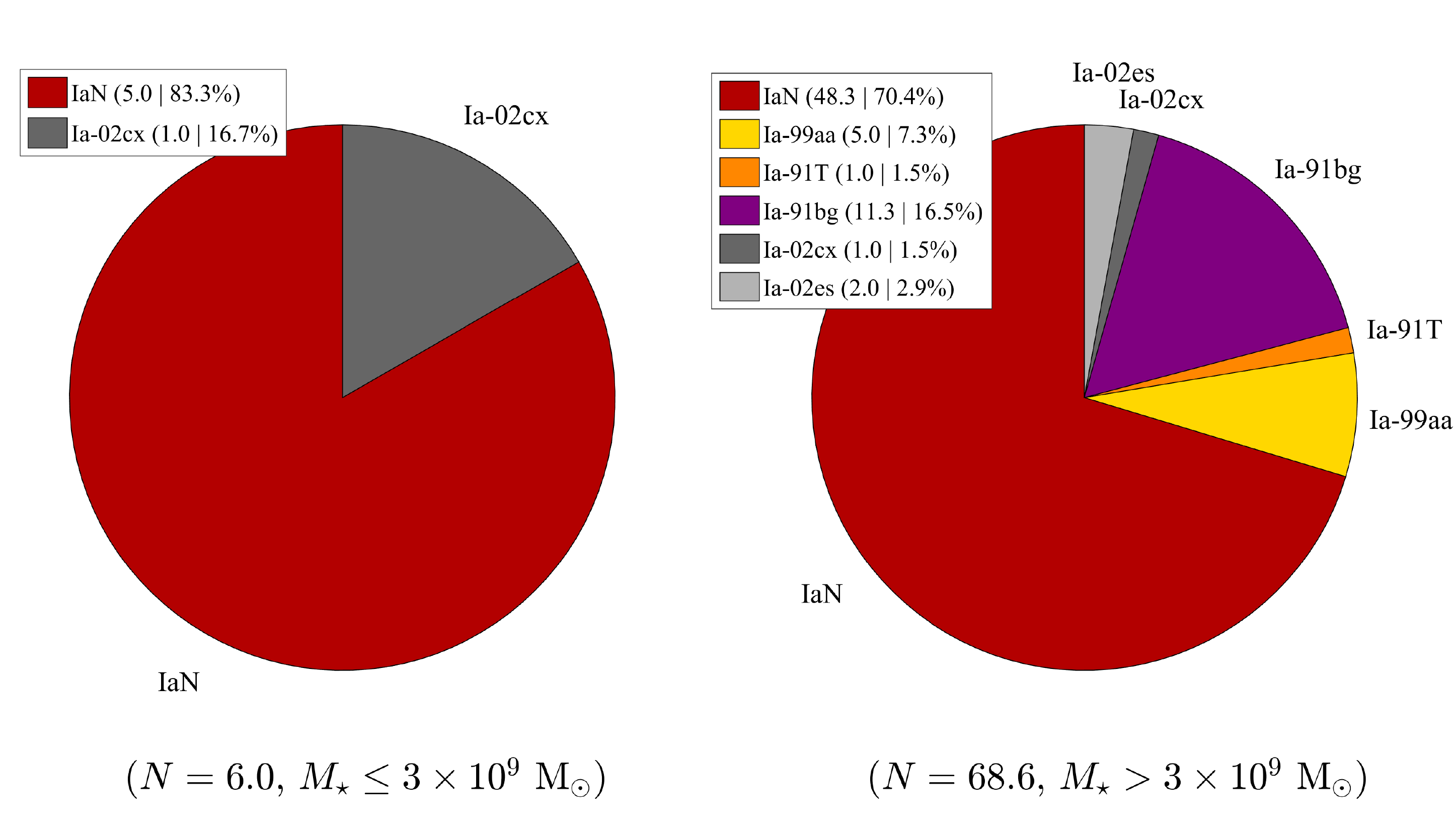}
 \caption{Relative SN rates in the LOSS volume-limited sample, with a mass cut at $3\times10^{9}~{\rm M}_\sun$ (the mass of the LMC, often used to define dwarf galaxies). Because the samples of SNe in low-mass galaxies are small, the only significant trend is an overrepresentation of SNe IIn in low-mass galaxies.}
 \label{fig:pie_3e9}
\end{figure*}

\begin{figure*}
 \center
 \plotone{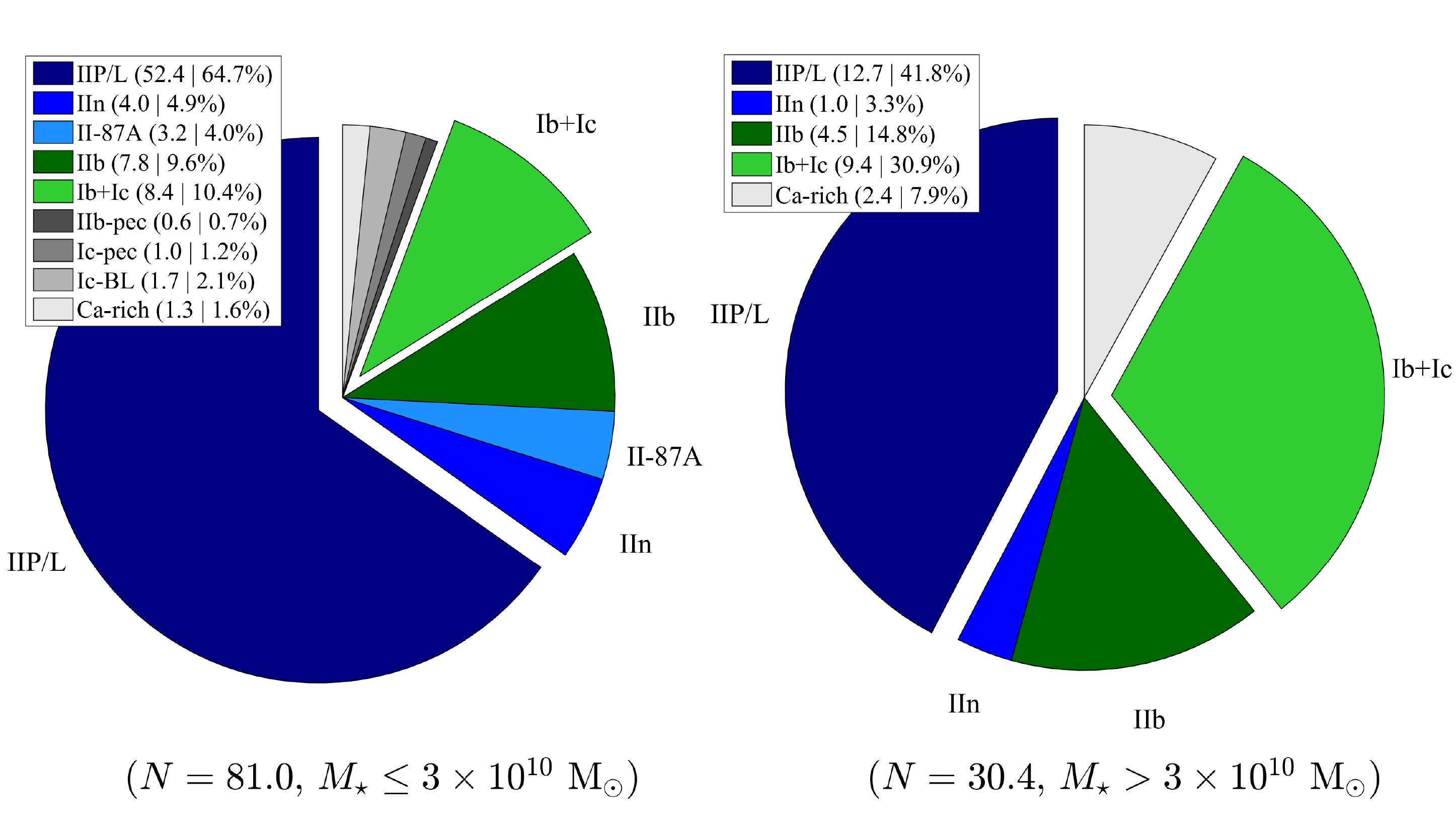} \\
 \plotone{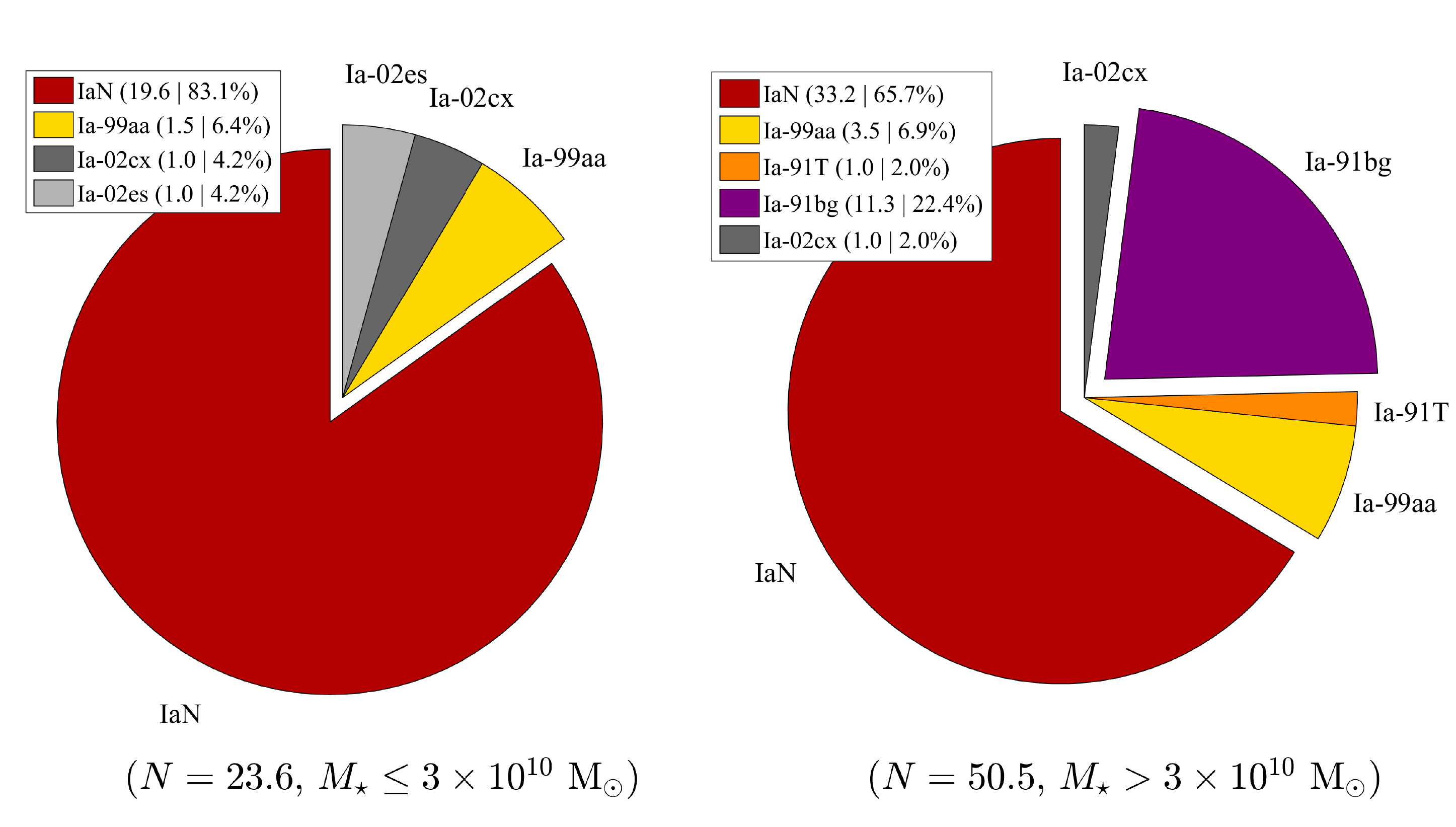}
 \caption{Relative SN rates in the LOSS volume-limited sample, with a mass cut at $3\times10^{10}~{\rm M}_\sun$. The same qualitative trends seen in Figure~\ref{fig:pie} are apparent here as well, but the statistical significance of some relative rates has changed, as shown by the exploded slices. SNe Ib and Ic are still significantly underrepresented in low-mass galaxies. SN 1987A-like SNe are still only found in low-mass galaxies, but this is no longer a significant result. SNe IIP/L, however, are now significantly overrepresented in low-mass galaxies. Normal SNe Ia are still significantly underrepresented in high-mass galaxies, where the overrepresentation of SN 1991bg-like SNe~Ia is now significant as well.}
 \label{fig:pie_3e10}
\end{figure*}

Here, we present the relative SN rates in low- and high-mass galaxies when splitting the galaxy sample at either $3\times10^9~{\rm M}_\sun$ (Table~\ref{table:fractions_mass3e9} and Figure~\ref{fig:pie_3e9}) or $3\times10^{10}~{\rm M}_\sun$ (Table~\ref{table:fractions_mass3} and Figure~\ref{fig:pie_3e10}). We find similar qualitative trends as those described in Section~\ref{sec:vol_lim}, but the statistical significance of some trends changes as a result of changes in the number of SNe in each galaxy sample. There are two main differences between the results shown here and in Section~\ref{sec:vol_lim}, namely: when split at $3\times10^{10}~{\rm M}_\sun$, SNe IIP/L are significantly overrepresented in low-mass galaxies, as are SNe IIn when the galaxy sample is split at $3\times10^9~{\rm M}_\sun$. As in both of these cases the number of SNe in either low- or high-mass galaxies is smaller than in Section~\ref{sec:vol_lim}, it remains to be seen whether these trends will also be revealed in a larger SN sample. The overabundance of normal SNe~Ia in low-mass galaxies, as well as the overabundance of SN 1991bg-like SNe~Ia in high-mass galaxies, remains constant in the figures below, and becomes statistically significant when the sample is split at $3\times10^{10}~{\rm M}_\sun$.

\floattable
\begin{deluxetable}{lCC|CC|c}
 \tablecaption{Relative SN Rates in the LOSS Sample; Galaxy Stellar Mass Cut $3\times10^{9}~{\rm M}_\sun$. \label{table:fractions_mass3e9}}
 \tablehead{
 \colhead{SN type} & \colhead{Number} & \colhead{Fraction} & \colhead{Number} & \colhead{Fraction} & \colhead{Probability}
 }
 \startdata
               & \multicolumn{2}{c}{$M_\star\leq 3\times10^{9}~{\rm M}_\sun$} & \multicolumn{2}{c}{$M_\star > 3\times10^{9}~{\rm M}_\sun$} & \\
  \multicolumn{6}{c}{All SNe, out to 60 Mpc} \\
  SNe Ia       & 5.0^{+2.0}_{-2.0} & 0.32^{+0.13}_{-0.13} & 32^{+6}_{-6}  & 0.24^{+0.04}_{-0.04} & 20\% \\ 
  SE SNe       & 3.2^{+2.0}_{-1.6} & 0.20^{+0.13}_{-0.10} & 35^{+6}_{-6}  & 0.33^{+0.10}_{-0.08} & 21\% \\
  SNe II       & 8.0^{+3.0}_{-3.0} & 0.48^{+0.20}_{-0.17} & 66^{+10}_{-9} & 0.50^{+0.07}_{-0.07} & 41\% \\
  Total        & \multicolumn{2}{c}{15.8}                    & \multicolumn{2}{c}{132.8}                    &      \\
  \multicolumn{6}{c}{CC SNe, out to 60 Mpc} \\
  IIP/L        & 4.6^{+2.6}_{-2.2} & 0.42^{+0.24}_{-0.20} & 61^{+9}_{-9}       & 0.60^{+0.09}_{-0.09} & 14\%  \\ 
  II-87A       & 1.1^{+1.1}_{-1.1} & 0.10^{+0.10}_{-0.10} & 2.0^{+1.0}_{-1.0}  & 0.02^{+0.01}_{-0.01} & 14\%  \\
  IIn          & 2.0^{+1.0}_{-2.0} & 0.18^{+0.09}_{-0.18} & 3.0^{+2.0}_{-2.0}  & 0.03^{+0.02}_{-0.02} & 2.9\% \\
  IIb          & 2.1^{+1.6}_{-1.1} & 0.19^{+0.15}_{-0.10} & 10.1^{+3.3}_{-3.0} & 0.10^{+0.03}_{-0.03} & 18\%  \\
  Ib           & 0.5^{+0.5}_{-0.5} & 0.05^{+0.05}_{-0.05} & 10.4^{+4.1}_{-3.5} & 0.103^{+0.041}_{-0.035} & 49\%  \\
  Ic           & 0.5^{+0.5}_{-0.5} & 0.05^{+0.05}_{-0.05} & 6.3^{+2.5}_{-2.2}  & 0.063^{+0.025}_{-0.022} & 67\%  \\ 
  (Ib$+$Ic)\tablenotemark{a} & 1.1^{+0.5}_{-0.5}  & 0.10^{+0.05}_{-0.05} & 17^{+5}_{-4} & 0.17^{+0.05}_{-0.04} & 26\% \\
  IIb-pec      & 0.0^{+1.8}_{-0.0}  & 0.000^{+0.17}_{-0.00} & 0.6^{+0.6}_{-0.6}  & 0.006^{+0.006}_{-0.006} & 70\%  \\
  Ic-pec       & 0.0^{+1.8}_{-0.0}  & 0.000^{+0.17}_{-0.00} & 1.0^{+1.0}_{-1.0}  & 0.010^{+0.010}_{-0.010} & 70\%  \\
  Ic-BL        & 0.0^{+1.8}_{-0.0}  & 0.000^{+0.17}_{-0.00} & 1.7^{+1.3}_{-1.0}  & 0.017^{+0.013}_{-0.010} & 58\%  \\
  Ca-rich      & 0.0^{+1.8}_{-0.0}  & 0.000^{+0.17}_{-0.00} & 3.7^{+2.4}_{-2.3}  & 0.037^{+0.024}_{-0.023} & 44\%  \\
  Total        & \multicolumn{2}{c}{10.9}                     & \multicolumn{2}{c}{100.6}                    &       \\
  \multicolumn{6}{c}{SNe Ia, out to 80 Mpc} \\
  IaN          & 5.0^{+2.0}_{-2.0}  & 0.8^{+0.3}_{-0.3}    & 48^{+7}_{-7}       & 0.7^{+0.1}_{-0.1}       & 26\%  \\
  Ia-99aa      & 0.0^{+1.8}_{-0.0}  & 0.0^{+0.3}_{-0.3}    & 5.0^{+2.0}_{-2.0}  & 0.07^{+0.03}_{-0.03}    & 44\%  \\
  Ia-91T       & 0.0^{+1.8}_{-0.0}  & 0.0^{+0.3}_{-0.3}    & 1.0^{+1.0}_{-1.0}  & 0.015^{+0.015}_{-0.015} & 74\%  \\
  Ia-91bg      & 0.0^{+1.8}_{-0.0}  & 0.0^{+0.3}_{-0.3}    & 11^{+3}_{-3}       & 0.165^{+0.047}_{-0.045} & 22\% \\
  Ia-02cx      & 1.0^{+1.0}_{-1.0}  & 0.17^{+0.17}_{-0.17} & 1.0^{+1.0}_{-1.0}  & 0.015^{+0.015}_{-0.015} & 74\%  \\
  Ia-02es      & 0.0^{+1.8}_{-0.0}  & 0.0^{+0.3}_{-0.3}    & 2.0^{+1.0}_{-1.0}  & 0.029^{+0.015}_{-0.015} & 64\%  \\
  Total        & \multicolumn{2}{c}{6.0}                      & \multicolumn{2}{c}{68.6}                     &       \\
 \enddata   
 \tablenotetext{a}{Because the classification of some SNe as either SNe Ib or Ic is uncertain, we also present the fraction of both types, combined.}
\end{deluxetable}

\floattable
\begin{deluxetable}{lCC|CC|c}
 \tablecaption{Relative SN Rates in the LOSS Sample; Galaxy Stellar Mass Cut $3\times10^{10}~{\rm M}_\sun$. \label{table:fractions_mass3}}
 \tablehead{
 \colhead{SN type} & \colhead{Number} & \colhead{Fraction} & \colhead{Number} & \colhead{Fraction} & \colhead{Probability}
 }
 \startdata
               & \multicolumn{2}{c}{$M_\star\leq 3\times10^{10}~{\rm M}_\sun$} & \multicolumn{2}{c}{$M_\star > 3\times10^{10}~{\rm M}_\sun$} & \\
  \multicolumn{6}{c}{All SNe, out to 60 Mpc} \\
  SNe Ia       & 16.6^{+4.5}_{-4.0} & 0.17^{+0.05}_{-0.04} & 20^{+5}_{-4}       & 0.40^{+0.10}_{-0.08} & $<0.1$\% \\ 
  SE SNe       & 21^{+4}_{-4}       & 0.22^{+0.05}_{-0.04} & 16^{+5}_{-4}       & 0.33^{+0.10}_{-0.08} & 4.9\% \\
  SNe II       & 60^{+9}_{-9}       & 0.61^{+0.09}_{-0.09} & 13.8^{+3.9}_{-3.6} & 0.27^{+0.08}_{-0.07} & $<0.1$\% \\
  Total        & \multicolumn{2}{c}{97.3}                     & \multicolumn{2}{c}{50.5}                     &       \\
  \multicolumn{6}{c}{CC SNe, out to 60 Mpc} \\
  IIP/L        & 52^{+9}_{-9}       & 0.65^{+0.110}_{-0.11}   & 12.7^{+4.0}_{-3.5} & 0.42^{+0.13}_{-0.12} & 1.2\% \\ 
  II-87A       & 3.2^{+2.1}_{-2.0}  & 0.040^{+0.026}_{-0.025} & 0.0^{+1.8}_{-0.0}  & 0.00^{+0.06}_{-0.00} & 19\% \\
  IIn          & 4.0^{+2.0}_{-2.0}  & 0.049^{+0.025}_{-0.025} & 1.0^{+1.0}_{-1.0}  & 0.03^{+0.03}_{-0.03} & 33\%  \\
  IIb          & 7.8^{+2.9}_{-2.6}  & 0.096^{+0.036}_{-0.032} & 4.5^{+2.1}_{-2.0}  & 0.15^{+0.07}_{-0.07} & 20\%  \\
  Ib           & 5.0^{+2.5}_{-2.0}  & 0.062^{+0.031}_{-0.025} & 5.8^{+3.5}_{-3.0}  & 0.19^{+0.12}_{-0.10} & 1.2\% \\
  Ic           & 3.2^{+1.7}_{-1.4}  & 0.040^{+0.021}_{-0.017} & 3.6^{+2.0}_{-1.6}  & 0.12^{+0.07}_{-0.05} & 2.8\% \\ 
  (Ib$+$Ic)\tablenotemark{a} & 8.4^{+3.0}_{-2.6}  & 0.10^{+0.04}_{-0.03} & 9.4^{+4.1}_{-3.5} & 0.31^{+0.14}_{-0.12} & 0.3\% \\
  IIb-pec      & 0.6^{+0.6}_{-0.6}  & 0.007^{+0.007}_{-0.007} & 0.0^{+1.8}_{-0.0}  & 0.00^{+0.06}_{-0.00} & 44\%  \\
  Ic-pec       & 1.0^{+1.0}_{-1.0}  & 0.012^{+0.012}_{-0.012} & 0.0^{+1.8}_{-0.0}  & 0.00^{+0.06}_{-0.06} & 44\% \\
  Ic-BL        & 1.7^{+1.3}_{-1.0}  & 0.021^{+0.016}_{-0.012} & 0.0^{+1.8}_{-0.0}  & 0.00^{+0.06}_{-0.00} & 28\%  \\
  Ca-rich      & 1.3^{+1.3}_{-1.3}  & 0.016^{+0.016}_{-0.016} & 2.4^{+1.4}_{-1.3}  & 0.08^{+0.05}_{-0.04} & 6.7\% \\
  Total        & \multicolumn{2}{c}{81.0}                     & \multicolumn{2}{c}{30.4}                     &       \\
  \multicolumn{6}{c}{SNe Ia, out to 80 Mpc} \\
  IaN          & 19.6^{+4.5}_{-4.5} & 0.8^{+0.2}_{-0.2}       & 33^{+6}_{-6}       & 0.66^{+0.12}_{-0.11} & 3.5\%  \\
  Ia-99aa      & 1.5^{+1.0}_{-1.0}  & 0.06^{+0.04}_{-0.04}    & 3.5^{+2.0}_{-1.5}  & 0.07^{+0.04}_{-0.03} & 57\%  \\
  Ia-91T       & 0.0^{+1.8}_{-0.0}  & 0.00^{+0.08}_{-0.00}    & 1.0^{+1.0}_{-1.0}  & 0.02^{+0.02}_{-0.02} & 40\%  \\
  Ia-91bg      & 0.0^{+1.8}_{-0.0}  & 0.00^{+0.08}_{-0.00}    & 11^{+3}_{-3}       & 0.22^{+0.06}_{-0.06} & 0.3\% \\
  Ia-02cx      & 1.0^{+1.0}_{-1.0}  & 0.04^{+0.04}_{-0.04}    & 1.0^{+1.0}_{-1.0}  & 0.02^{+0.02}_{-0.02} & 29\%  \\
  Ia-02es      & 1.0^{+2.0}_{-1.0}  & 0.04^{+0.09}_{-0.04}    & 0.0^{+1.0}_{-0.0}  & 0.00^{+0.02}_{-0.00} & 14\%  \\
  Total        & \multicolumn{2}{c}{23.6}                     & \multicolumn{2}{c}{50.5}                     &       \\
 \enddata   
 \tablenotetext{a}{Because the classification of some SNe as either SNe Ib or Ic is uncertain, we also present the fraction of both types, combined.}
\end{deluxetable}

\end{document}